\documentclass[epj]{svjour}
\usepackage{graphics,bbm,amssymb}
\begin{document}
%
\title{Controlling qubit arrays with anisotropic $XXZ$ 
Heisenberg interaction by acting on a single qubit}
\author{Rahel Heule\inst{1} \and C. Bruder\inst{1} \and Daniel Burgarth\inst{2,3}
\and Vladimir M. Stojanovi\'c\inst{1}}                     
\institute{Department of Physics, University of Basel,
 Klingelbergstrasse 82, CH-4056 Basel, Switzerland 
\and Institute for Mathematical Sciences, Imperial College London, 
SW7 2PG, United Kingdom 
\and QOLS, The Blackett Laboratory, Imperial College London, 
Prince Consort Road, SW7 2BW, United Kingdom}
\date{Received: date / Revised version: date}
%

\abstract{We investigate anisotropic $XXZ$ Heisenberg spin-$1/2$
  chains with control fields acting on one of the end spins, with the
  aim of exploring local quantum control in arrays of interacting
  qubits. In this work, which uses a recent Lie-algebraic result on
  the local controllability of spin chains with ``always-on''
  interactions, we determine piecewise-constant control pulses
  corresponding to optimal fidelities for quantum gates such as
  spin-flip (NOT), controlled-NOT (CNOT), and square-root-of-SWAP
  ($\sqrt{\textrm{SWAP}}$).  We find the minimal times for realizing
  different gates depending on the anisotropy parameter $\Delta$ of
  the model, showing that the shortest among these gate times are
  achieved for particular values of $\Delta$ larger than unity. To study
  the influence of possible imperfections in anticipated experimental
  realizations of qubit arrays, we analyze the robustness of the
  obtained results for the gate fidelities to random variations in the
  control-field amplitudes and finite rise time of the
  pulses. Finally, we discuss the implications of our study for
  superconducting charge-qubit arrays.
} 
\maketitle
\section{Introduction}\label{intro}
Coherent control of quantum systems is one of the prerequisites for
quantum information processing. While already simple arguments lead to
the conclusion that almost any coupled quantum system can be
controlled in principle~\cite{Lloyd+:04}, the mathematical foundations
of the subject are based on the notion of controllability and
formulated using the language of Lie algebras~\cite{D'AlessandroBook}.
In particular, a system is completely controllable if its internal dynamics
governed by external fields can give rise to an arbitrary unitary
transformation in the Hilbert space of the
system~\cite{Jurdjevic+Sussmann:72}.  Both state control and the more
general operator control have been implemented in a variety of
systems~\cite{ControlRevEPJD}.

Recent quantum control studies have focused their attention on
interacting systems.  A familiar example is furnished by spin chains,
systems that can be used as data buses~\cite{SBose:07} for
state-~\cite{Romito+:05,Lyakhov+Bruder:06,Burgarth:07,Caneva++:09} and
entanglement transfer~\cite{Maruyama+:07}. In such systems,
``always-on'' interactions between the constituents (typically nearest
neighbors) allow for a global control of the system dynamics by
manipulating only a small subsystem, in the extreme case a single
spin. The main question is then what is the smallest possible
subsystem of a given system that one needs to act upon to
ensure the complete controllability, or, at least, the ability to
perform certain pre-determined unitary transformations. This is the
central idea behind the {\em local-control} approach.

The fact that the local-control approach can be advantageous in
interacting systems provides an incentive for identifying minimal
controlling resources that guarantee controllability in particular
classes of systems. Quite recently, several Lie-algebraic results
pertaining to local control of spin chains have been
obtained~\cite{Schirmer:08,Burgarth+:09,Kay+Pemberton:10,Burgarth+:10,Wang+:10}.
For example, it was demonstrated that acting only on one of the end
spins of an $XXZ$-Heisenberg spin chain ensures complete
controllability of the chain~\cite{Burgarth+:09}.  Adopting the last
result as our point of departure, in this paper we investigate the
feasibility of local operator control in qubit arrays modeled as
spin-$1/2$ chains with Heisenberg interaction.  In contrast to our
recent proof-of-principle study~\cite{Heule++:10}, where only the
isotropic Heisenberg-coupling case was addressed, in the present work
we discuss the case of (anisotropic) $XXZ$ coupling. The main
motivation stems from the relevance of the $XXZ$-case for
implementations of Josephson-junction based superconducting qubit
arrays~\cite{Makhlin+:01,Levitov++:01,You+Nori:05}.

We determine piecewise-constant control fields, acting only on the
first spin in the chain, which lead to the highest possible fidelities
for a selected set of quantum logic operations: the spin-flip (NOT) of
the last spin in the chain, as well as the controlled-NOT (CNOT) and
the square-root-of-$\textrm{SWAP}$ ($\sqrt{\textrm{SWAP}}$) gates
applied to the last two spins. We optimize the gate fidelities with
respect to the control-field amplitudes for three-spin chains.  
We then carry out a sensitivity analysis, i.e., discuss the
robustness of the obtained results with respect to random errors in
the control fields, as well as finite rise/decay-times for control-field
amplitudes. The present work is concerned with $XXZ$-Heisenberg spin
chains and our conclusions apply to any physical realization of qubit
arrays with this type of coupling~\cite{Makhlin+:01,Levitov++:01}.

\section{System and method} \label{system}
The total Hamiltonian of a Heisenberg spin-$1/2$ chain of length $N_{s}$ reads 
\begin{equation}
H(t)=H_{0}+ H_c(t)\:,
\end{equation}
\noindent where
\begin{equation}\label{Heisenberg}
H_{0}=J\sum_{i=1}^{N_{s}-1}\:\left(S_{i,x}S_{i+1,x}+S_{i,y}S_{i+1,y}
+\Delta S_{i,z}S_{i+1,z}\right) \:,
\end{equation}
\noindent is a $XXZ$ Heisenberg part with anisotropy $\Delta$, and    
\begin{equation}\label{controlham}
H_c(t)=h_x(t)S_{1x}+h_y(t)S_{1y}
\end{equation}
\noindent a Zeeman-like control part, with control fields $h_x(t)$,
$h_y(t)$ acting only on the first spin. In what follows, we will also
employ control Hamiltonians with fields in the $x$ and
$z$-directions. Whether the $XXZ$ spin chain
under consideration is ferromagnetic or antiferromagnetic is not
crucial here, as we are concerned with operator control; aspects such
as, for example, the different nature of the ground states in the two
cases (separable vs. entangled) would only be consequential for issues
related to, e.g., state control or entanglement transfer. For
definiteness, we will assume that $J>0$ and $\Delta>0$. 
It is useful to recall that the one-dimensional $XXZ$
model has an antiferromagnetic ground state for $\Delta\geq 1$, a
ferromagnetic one for $\Delta< -1$, while for the intermediate values
of $\Delta$ it is characterized by a critical gapless
(quasi-long-range ordered) phase~\cite{GiamarchiBook}. 

For convenience, we hereafter set $\hbar=1$ and, in addition, express
all frequencies and control fields in units of the coupling strength
$J$. Consequently, all times in the problem are expressed in units of
$1/J$.

Since implementing control fields with a complex time dependence
is difficult, we resort to piecewise-constant ones according to the
following scheme. At $t=0$ we start acting on the first spin of the
chain with an $x$ control pulse of amplitude $h_{x,1}$, which is
kept constant until $t=T$. Thus the system is governed by the
Hamiltonian $H_{x,1}\equiv H_0+h_{x,1}S_{1x}$. We then apply a $y$
pulse with the amplitude $h_{y,1}$ (Hamiltonian $H_{y,1}\equiv
H_0+h_{y,1}S_{1y}$) over the next interval of length $T$, etc. This
sequence repeats until $N_t$ pulses are carried out at $t=t_f\equiv
N_tT$. The full time evolution is described by
\begin{equation}\label{time_evolution_xy}
U(t_f)=U_{y,N_t/2}\:U_{x,N_t/2}\:\ldots\:U_{y,1}\:U_{x,1}\:,
\end{equation}
where $U_{x,i}\equiv e^{-iH_{x,i}T}$ and $U_{y,i}\equiv
e^{-iH_{y,i}T}$ are the respective time-evolution operators corresponding to 
$H_{x,i}$ and $H_{y,i}$, which can be evaluated using their spectral form. 

Our control objectives (target unitary operations) are both one-qubit
gates, such as the spin-flip ($\mathrm{NOT}$) operation on the last
spin of the chain
$X_{N_s}:=\mathbbm{1}\otimes\mathbbm{1}\otimes\ldots\otimes\mathbbm{1}\otimes
X$ ($X$ being the Pauli matrix), and some entangling two-qubit
gates. For instance,
$\mathrm{CNOT}_{N_s}:=
\mathbbm{1}\otimes\mathbbm{1}\otimes\ldots\otimes\mathbbm{1}\otimes\mathrm{CNOT}$
performs the controlled-NOT operation on the last two qubits in the
chain.  Similarly,
$\sqrt{\textrm{SWAP}}_{N_s}:=\mathbbm{1}\otimes\mathbbm{1}\otimes\ldots
\otimes\mathbbm{1}\otimes\sqrt{\textrm{SWAP}}$ performs the
$\sqrt{\textrm{SWAP}}$ operation on the same pair of qubits.

Unlike in many other control studies~\cite{Schirmer:08}, which make
use of single-excitation subspaces, we retain the full Hilbert space
of the system. This puts constraints on the system size that can be
treated within our framework. In what follows, we discuss three-spin
chains.

\section{Controllability and reachability} \label{controllability}
In Ref.~\cite{Burgarth+:09} a very general graph-infection criterion
was proven, which -- as a special case -- guarantees the complete
controllability of $XXZ$ Heisenberg spin chains through acting on one
end spin. The more conventional approaches for proving
complete controllability entail finding the dimension of the relevant
dynamical Lie algebra, a task for which special algorithms have been
developed~\cite{Schirmer+:01}. In the present problem, such an
algebra is generated by the skew-Hermitian traceless operators
$\{-iH_0,-iS_{1x},-iS_{1y}\}$ and has dimension $d^2-1$, where
$d\equiv 2^{N_s}$ is the dimension of the Hilbert space of the system.
Being generated by traceless operators, this algebra is then
isomorphic to $su(d)$, the Lie algebra associated with the special
unitary group $SU(d)$~\cite{PfeiferBook}.

Setting aside the issue of complete controllability, one might be
interested to know if some particular unitary operations -- on an
otherwise not completely controllable system~\cite{Polack+:09,Sander+Schulte-Herbrueggen:09} -- are
possible with an even smaller degree of manipulation, e.g.., a control
field only in one direction. For such operations,
equation~(\ref{time_evolution_xy}) goes over into
$U(t_f)=U_{x,N_t}\:\ldots\:U_{x,1}$. For example, the $X_{N_s}$ and
$\sqrt{\textrm{SWAP}}_{N_s}$ gates require only a control field in the
$x$ direction. To demonstrate this for $X_{N_s}$, let $\mathcal{L}_x$
be the dynamical Lie algebra generated by $-iH_0$ and $-iS_{1x}$, a
subalgebra of $su(d)$ with dimension $30$ in a three-spin $XXZ$ chain
(note that the counterpart of this algebra in the isotropic-coupling
case has smaller dimension, namely $18$). For showing that $X_{N_s}$
belongs to the connected Lie subgroup $e^{\mathcal{L}_x}$ of $SU(d)$
it suffices to find an element $A\in\mathcal{L}_x$ such that
$X_{N_s}=e^{A}$. Using the repeated commutators of the generators of
$\mathcal{L}_x$, it can be demonstrated that $X_{N_s}$ is an element
of this algebra. $X_{N_s}$ is both unitary and Hermitian, implying
that $X_{N_s}^2=\mathbbm{1}$. It is then easy to show that
$A=-i\frac{\pi}{2}X_{N_s}$, an element of $\mathcal{L}_x$, fulfills
$e^{A}=-iX_{N_s}$. Therefore, $X_{N_s}$ is reachable using only an
$x$ control field. Recalling that the $\sqrt{\textrm{SWAP}}$ gate on
two qubits is given by~\cite{Burkard++:99}
\begin{equation}
\sqrt{\textrm{SWAP}}=e^{i\frac{\pi}{8}}
e^{-i\frac{\pi}{8}(X\otimes X+Y\otimes Y+Z\otimes Z)}\: ,
\end{equation}
the reachability of $\sqrt{\textrm{SWAP}}_{N_s}$ using an $x$ control only readily follows from
the fact that 
$\mathbbm{1}\otimes\mathbbm{1}\otimes\ldots\otimes\mathbbm{1}\otimes
(X\otimes X+Y\otimes Y+Z\otimes Z)$ is an element of $\mathcal{L}_x$.

\section{Target gates and minimal gate times}\label{targetgates}
In this section our goal is to find control fields leading to optimal
fidelities for a chosen set of quantum gates, with a particular emphasis 
on minimal times needed for realizing different gates depending on the 
anisotropy $\Delta$.  

In quantum operator control, the figure of merit is the gate fidelity
\begin{equation}\label{deffidelity}
F(t_{f})=\frac{1}{d}\big|\mathrm{tr}
\big[U^{\dag}(t_{f})U_{\mathrm{target}}\big]\big|\:,
\end{equation}
where $U(t_{f})$ is the time-evolution operator of the system at time
$t=t_{f}$ (Eq.~(\ref{time_evolution_xy})) and $U_{\mathrm{target}}$
stands for the quantum gate that we want to realize. We perform
optimization, i.e., maximize the gate fidelity with respect to the
$N_{t}$ control-field amplitudes, for varying number ($N_{t}$) and
durations ($T$) of pulses (hence different total evolution times
$t_{f}$). We make use of a quasi-Newton method due to Broyden,
Fletcher, Goldfarb, and Shanno (BFGS-algorithm)~\cite{NRfortranBook}.
It should be stressed that, much like other optimization approaches,
this algorithm ensures only convergence to a local maximum. Therefore,
to determine a globally-optimal sequence of control-field amplitudes
(for a given target gate and given value of $\Delta$) we ought to
repeat the optimization process for a number of different initial
guesses for these amplitudes. We generate these initial guesses using
a uniform random number generator~\cite{NRfortranBook}.
\begin{figure}[t]
\resizebox{0.4\textwidth}{!}{\includegraphics{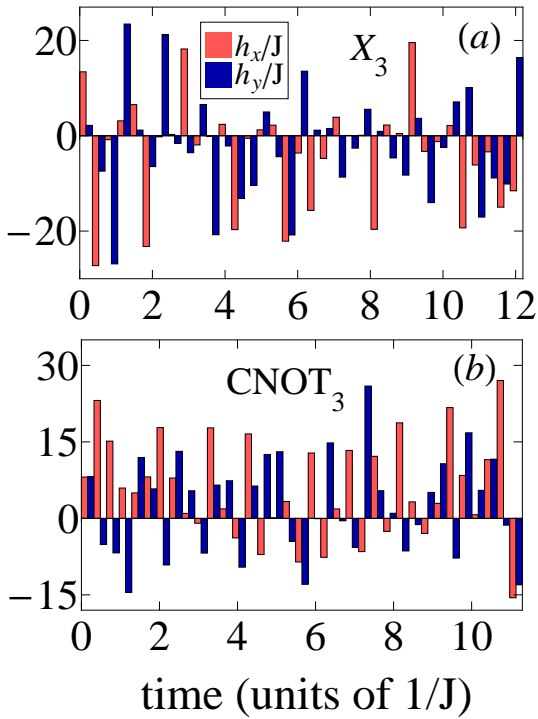}} 
\caption{Optimal control sequences for $\Delta=5$ realizing the 
(a) $X_3$ and (b) CNOT$_{3}$ gates with fidelity higher than 
0.999.} 
\label{ControlSeqFig}       
\end{figure}

An alternative to fixing the pulse durations and maximizing over the
control-field amplitudes would be to keep the amplitudes constant and
treat the pulse durations as variable control parameters. However, we
choose optimization over the control-field amplitudes since this
approach allows us to easily fix $t_{f}$ and determine its minimal value for
implementing the desired gate for any fixed value of the parameter
$\Delta$.
\begin{center}
\begin{table}[t]
\caption{Minimal times (in units of $J^{-1}$) needed to reach
  fidelities higher than $0.999$ for the relevant gates in the $x$-$y$ 
  control case. The corresponding values in the $x$-$z$ control
  case are given in the brackets.} \label{MinimalTimesTable1}
\begin{tabular}{@{}c@{\hspace{0.28cm}}c@{\hspace{0.28cm}}c@{\hspace{0.28cm}}c}
\hline\noalign{\smallskip}
$\Delta$ & X$_{3}$ & CNOT$_{3}$ & $\sqrt{\textrm{SWAP}}_{3}$ \\
\noalign{\smallskip}\hline\noalign{\smallskip}
\begin{tabular}{r@{.}l}
0&1\\
0&2\\
0&7\\
0&8\\
0&9\\
1&0\\
1&1\\
1&2\\
1&3\\
2&0\\
3&0\\
4&0\\
5&0\\
6&0\\
7&0\\
8&0\\
9&0\\
10&0\\
11&0\\
12&0\\
\end{tabular}
&
\begin{tabular}{r@{.}lr@{.}l}
23&3 & (19&0)\\
23&9 & (23&8)\\
18&2 & (18&0)\\
16&5 & (16&9)\\
15&4 & (13&9)\\
15&0 & (14&9)\\
15&4 & (15&8)\\
14&8 & (14&8)\\
16&2 & (16&0)\\
15&8 & (15&1)\\
12&0 & (13&4)\\
12&2 & (13&0)\\
12&2 & (13&2)\\
12&0 & (13&5)\\
12&9 & (13&9)\\
12&9 & (14&5)\\
13&6 & (15&0)\\
14&8 & (16&6)\\
21&7 & (17&9)\\
22&1 & (20&4)\\
\end{tabular}
&
\begin{tabular}{r@{.}lr@{.}l}
73&2 & (60&2)\\
33&5 & (28&7)\\
22&2 & (22&6)\\
22&8 & (21&6)\\
18&4 & (18&2)\\
17&3 & (16&7)\\
17&3 & (16&4)\\
20&6 & (16&7)\\
20&0 & (19&0)\\
12&3 & (12&4)\\
12&0 & (12&2)\\
12&2 & (12&2)\\
11&2 & (11&1)\\
11&5 & (11&3)\\
11&8 & (9&8)\\
13&3 & (11&7)\\
13&1 & (10&4)\\
14&4 & (11&9)\\
13&1 & (10&8)\\
19&3 & (12&0)\\
\end{tabular}
&
\begin{tabular}{r@{.}lr@{.}l}
14&2 & (14&2)\\
20&7 & (21&2)\\
18&9 & (16&8)\\
16&6 & (16&9)\\
1&6 & (1&6)\\
1&5 & (1&5)\\
1&5 & (1&5)\\
14&9 & (12&2)\\
15&2 & (14&8)\\
12&8 & (9&9)\\
10&2 & (8&8)\\
8&0 & (6&6)\\
5&4 & (5&3)\\
4&4 & (4&4)\\
3&8 & (2&0)\\
3&3 & (3&3)\\
4&3 & (3&0)\\
3&9 & (3&9)\\
2&4 & (2&4)\\
2&2 & (3&3)\\
\end{tabular}\\
\noalign{\smallskip}\hline
\end{tabular}
\vspace*{5cm}  
\end{table}
\end{center}
The obtained results for the gate fidelities have the following two
salient features.  Firstly, for fixed parameters of the model and
fixed total evolution time $t_{f}$, the fidelity for any given gate
can increase significantly with increasing $N_{t}$ (or, equivalently,
decreasing $T$). In other words, more rapid switching leads to higher
fidelities. For instance, in the case of the CNOT$_{3}$ gate with
$\Delta=1.3$ and $t_{f}=30$, for $N_{t}=10,20,30,40,50,60,70$ we
obtain the respective fidelities
$F=0.455,0.697,0.837,0.953,0.995,1-10^{-4},1-10^{-8}$.  Secondly, for
each gate there exists a minimal value of $t_{f}$ (i.e., minimal gate
time), below which fidelities close to unity cannot be reached
regardless of the value of $N_{t}$. The obtained minimal gate times
for different values of $\Delta$ in the $x$-$y$ ($x$-$z$) control
cases are given in Table~\ref{MinimalTimesTable1}.  Apparently, there
exists an optimal value of $\Delta$ which corresponds to the shortest
among these times. For the $X_{3}$ and CNOT$_{3}$ gates, for example,
these values are around $\Delta=5$. The corresponding optimal
sequences of $x$ and $y$ control pulses for the $X_{3}$ and CNOT$_{3}$
gates are shown in Figures~\ref{ControlSeqFig}(a) and
\ref{ControlSeqFig}(b), respectively. Since ideal steplike pulses
cannot be realized in practice, in Ref.~\cite{Heule++:10} we also
studied frequency-filtered control fields and showed that sufficiently
high fidelities can still be retained.

\begin{center}
\begin{table}[b]
\caption{Minimal times (in units of $J^{-1}$) needed to reach
  fidelities higher than $0.999$ for the relevant gates using 
  only an $x$ control field.} \label{MinimalTimesTable2}
\begin{tabular}{@{}c@{\hspace{0.28cm}}c@{\hspace{0.28cm}}c|@{}c@{\hspace{0.28cm}}c@{\hspace{0.28cm}}c}
\hline\noalign{\smallskip}
$\Delta$ & X$_{3}$ & $\sqrt{\textrm{SWAP}}_{3}$ & $\Delta$ & X$_{3}$ & $\sqrt{\textrm{SWAP}}_{3}$\\
\noalign{\smallskip}\hline\noalign{\smallskip}
\begin{tabular}{r@{.}l}
0&1\\
0&2\\
0&7\\
0&8\\
0&9\\
1&0\\
1&1\\
1&2\\
1&3\\
2&0\\
\end{tabular}
&
\begin{tabular}{r@{.}l}
128&4\\
70&4\\
36&4\\
37&1\\
34&4\\
18&8\\
25&2\\
31&1\\
30&7\\
23&2\\
\end{tabular}
&
\begin{tabular}{r@{.}l}
80&6\\
45&0\\
25&0\\
25&6\\
28&9\\
25&8\\
25&8\\
31&9\\
31&5\\
21&7\\
\end{tabular}
&
\begin{tabular}{r@{.}l}
3&0\\
4&0\\
5&0\\
6&0\\
7&0\\
8&0\\
9&0\\
10&0\\
11&0\\
12&0\\
\end{tabular}
&
\begin{tabular}{r@{.}l}
19&2\\
18&4\\
17&3\\
16&4\\
16&1\\
15&5\\
15&8\\
15&8\\
16&2\\
16&5\\
\end{tabular}
&
\begin{tabular}{r@{.}l}
14&2\\
9&5\\
7&7\\
7&5\\
7&3\\
6&4\\
5&7\\
3&9\\
4&7\\
4&3\\
\end{tabular}\\
\noalign{\smallskip}\hline
\end{tabular}
\vspace*{5cm}  
\end{table}
\end{center}

In Table~\ref{MinimalTimesTable2} the minimal times are given for the
$X_3$ and $\sqrt{\textrm{SWAP}}_{3}$ gates realized using only an $x$
control field.  It is interesting to compare these minimal times to
the above case with both $x$ and $y$ (or $x$ and $z$) controls. For
small values of $\Delta$ (with the exception of $\Delta=1$) the
minimal times for realizing the $X_3$ gate in the $x$-only control
case are significantly longer than their counterparts in the $x$-$y$
($x$-$z$) case.  In contrast, for larger $\Delta$ these times become
more and more similar.  Finally, for $\Delta\geq 11$ the minimal times
in the $x$-only control case are even shorter than in the $x$-$y$ and
$x$-$z$ cases. Since $x$-only control is easier to implement,
this surprising observation provides an additional argument for using
$x$-only control in the regime of interest for superconducting charge
qubits.

As is well known~\cite{Burkard++:99}, the $\sqrt{\textrm{SWAP}}$ gate
on two qubits is naturally implemented by the isotropic Heisenberg
Hamiltonian after a time $\tau=\pi/2\approx 1.57$. As can be seen in
Table~\ref{MinimalTimesTable1}, the minimal
$\sqrt{\textrm{SWAP}}_{3}$-gate times indeed seem to correspond to
$\Delta\approx 1$ and are only slightly shorter than in the
control-free case. This is despite the fact that our
$\sqrt{\textrm{SWAP}}_{3}$ gate performs the $\sqrt{\textrm{SWAP}}$
operation on the last two qubits (leaving the state of the first qubit
unchanged) while the Heisenberg Hamiltonian of
equation~(\ref{Heisenberg}) also contains the interaction between the
first two qubits. Thus we can conclude that the role of control fields
in this case is to counteract the effect of the free evolution of the
first qubit governed by $H_0$.

Generally speaking, the minimal gate times can in principle be found
based on the time-optimal unitary operation formalism put forward by
Carlini and co-workers~\cite{Carlini+:07}. This method requires
solving a system of coupled nonlinear equations for Lagrange
multipliers resulting from the quantum brachistochrone equation. In
practice, extracting minimal times for different quantum gates in this
way is feasible only when the time evolution of the total Hamiltonian
of the system is as simple as to allow for an analytical solution of
these equations. This is possible, for instance, when this Hamiltonian
has a block-diagonal form in the computational basis, where each block
commutes with itself at different times. In the problem at hand this
is not the case, therefore an alternative strategy for finding minimal
times is required.

\section{Robustness to random errors and finite pulse rise times}
\label{sensitivity}

In the following, we analyze the sensitivity of the fidelity to random
errors in the control-field amplitudes, as well as to a finite rise
time.

The random errors in control-field amplitudes are assumed to follow a
uniform distribution of half-width $\delta$. For given $\delta$, we
generate a large sample of $N\sim 1000$ control fields affected by
random noise, for which we recalculate the fidelity. We are interested
in the behavior of the average fidelity $\bar{F}=\sum_{i=1}^N F_i/N$,
where the $F_i$ are fidelities for specific realizations of the random
field, versus $\delta$ for the gates of interest and varying values of
$\Delta$.
\begin{figure}[t]
\resizebox{0.5\textwidth}{!}{\includegraphics{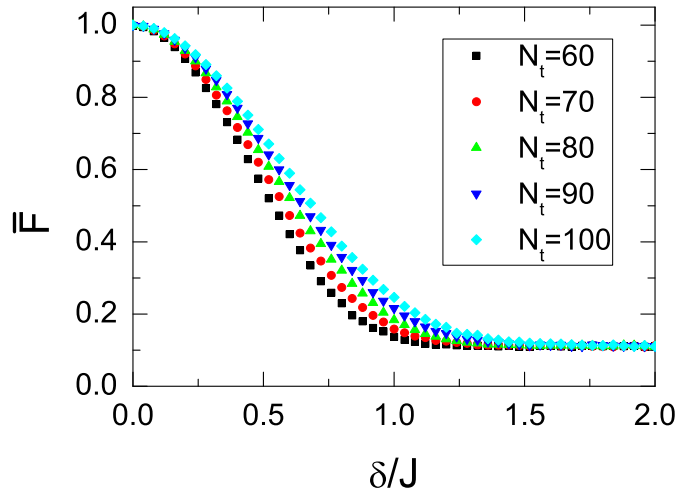}}
\caption{(Color online) Average fidelity versus half-width ($\delta$) 
for the $\sqrt{\textrm{SWAP}}_{3}$ gate with $\Delta=1.2$ and fixed total 
time $t_{f}=60$.} \label{SensitivFig2}       
\end{figure}

In our previous work~\cite{Heule++:10}, using the isotropic Heisenberg
model ($\Delta=1$) as an example, it was demonstrated that the shape
of the fidelity decay curves ($\bar{F}$ vs. $\delta$) depends on the
number of control pulses $N_t$ and their length $T$. Provided that the system satisfies the
conditions for complete controllability, the saturation regime of the
average fidelity sets in for $\delta\gtrsim 2J$. The universal saturation
value is $1/d$, where $d$ is the dimension of the Hilbert space of the
system. Importantly, for fixed $t_{f}=N_{t}T$, the average fidelity is
closer to the intrinsic (in the absence of random errors) optimal
values for larger $N_{t}$ (faster switching), this being a consequence
of general properties of systems that exhibit competition between the
resonance- and relaxation-type behavior~\cite{Heule++:10}.  Therefore,
more rapid switching leads not only to higher intrinsic fidelities in
the absence of randomness (recall section \ref{targetgates}), but also
renders these fidelities less sensitive to random errors. This is a
manifestation of an intrinsic robustness of the system.
Figure~\ref{SensitivFig2} illustrates that these features are also
present in the anisotropic XXZ case.
\begin{figure}[b]
\resizebox{0.5\textwidth}{!}{\includegraphics{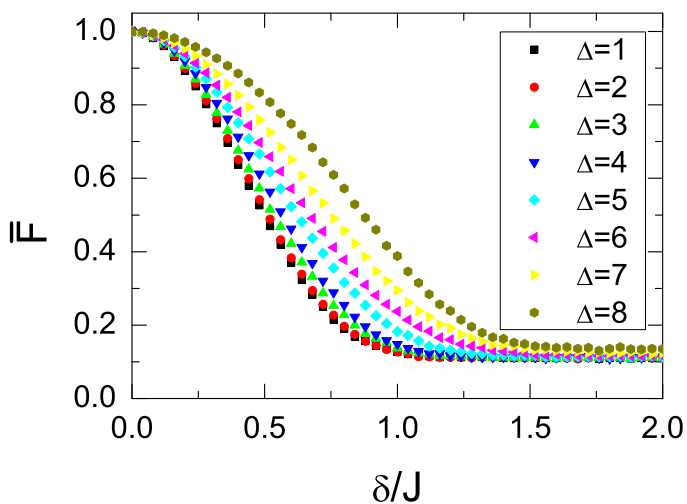}}
\caption{(Color online) Average fidelity versus half-width ($\delta$) 
of random-noise distribution for optimal control sequences with $N_{t}=70$ 
and $T=1$ corresponding to the CNOT$_{3}$ gate.} \label{SensitivityFig}       
\end{figure}

The sensitivity to random errors in the control-field amplitudes
depending on the anisotropy $\Delta$ is illustrated in 
Figure~\ref{SensitivityFig}. As can be inferred from this figure, for larger
$\Delta$ the system is less sensitive to random errors.

\begin{figure}[b]
\resizebox{0.5\textwidth}{!}{\includegraphics{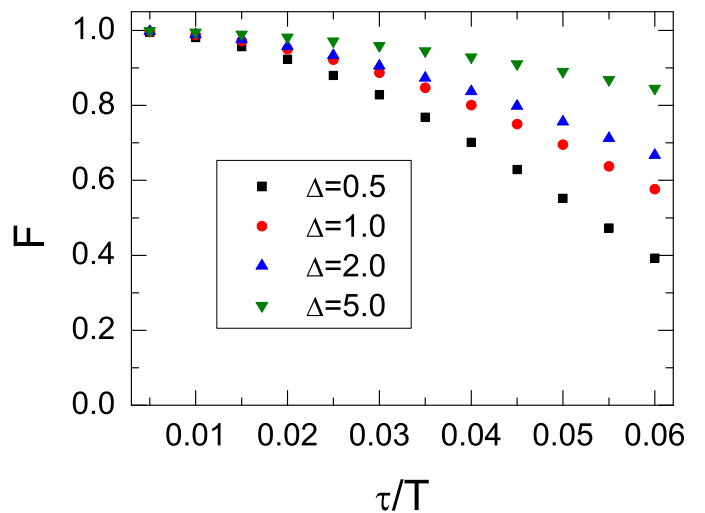}}
\caption{(Color online) Illustration of sensitivity to finite rise time for 
the $X_{3}$ gate. The optimal control sequences used correspond to $N_{t}=70$ 
and $T=1$.} \label{FiniteRiseFig}       
\end{figure}

Another unavoidable source of imperfections in qubit-array
realizations is the finite rise time of the control fields. Instead of
a stepwise behavior, experimental control
fields $h_{j,n}$ ($j=x,y;\:n=1,\ldots,N_{t}/2$) are expected to have a 
finite rise/decay time $\tau$. 
Figure~\ref{FiniteRiseFig} shows the dependence of the fidelity on the 
finite rise time. For larger values of $\Delta$, the fidelities of optimal 
control sequences seem to be less affected by the finite rise time.

The central result of this section is that values of $\Delta>1$ lead
to both shorter gate times and a reduced sensitivity of the fidelity
to random errors in control field and finite rise time.
This provides a guiding principle for future implementations of qubit arrays.

\section{Discussion and Conclusions}\label{conclude}
Our results are of direct relevance to superconducting qubit
arrays~\cite{Romito+:05}. One-dimensional Josephson arrays of
capacitively coupled superconducting islands can be described as $XXZ$
Heisenberg spin-$1/2$ chains~\cite{bruder++:93,fazio+:01}. In general,
the $XY$-part of Hamiltonian is characterized by a nearest-neighbor
interaction, whereas the $Z$-part will also have coupling
contributions beyond nearest neighbors. However, by properly choosing
the junction capacitances and the capacitance of each island to the
back gate of the structure, the $Z$-part will also be approximately of
nearest-neighbor type.  The correspondence between the parameters of
the Josephson array and the spin chain is as follows: the Josephson
energy $E_J$ of the junctions coupling the islands corresponds to the
exchange coupling constant $J$ of the spin system and can be
controlled by a magnetic field if we assume that the coupling
junctions are realized as SQUIDs. The parameter $\Delta J$ of the spin
system corresponds to the charging energy $E_C$, i.e., the anisotropy
parameter $\Delta$ corresponds to $E_C/E_J$. Values of $\Delta$ like
those studied in Tables~\ref{MinimalTimesTable1} and \ref{MinimalTimesTable2} 
can be experimentally realized. Finally, the first island should form a charge 
qubit, and the control field $h_z$ corresponds to the gate voltage, while $h_x$
and $h_y$ play the role of the Josephson energy of this charge qubit.
Our study shows that, in principle, arbitrary quantum algorithms can be
realized on a one-dimensional Josephson array by controlling only the
first island in the array.

In summary, we have shown that local control of the first spin of an
anisotropic $XXZ$ Heisenberg spin-$1/2$ chain enables universal
quantum computation.  Using a recent Lie-algebraic result on the local
controllability of spin chains with ``always-on'' interactions, we
have determined control pulses leading to optimal
fidelities for quantum gates such as spin-flip (NOT), controlled-NOT
(CNOT), and square-root-of-SWAP ($\sqrt{\textrm{SWAP}}$).  We have
found the minimal times for realizing different gates depending on the
anisotropy parameter $\Delta$ of the model, showing that the shortest
among these gate times are achieved for particular values of $\Delta$
larger than unity.  Another surprising result was that in the regime
of interest for superconducting charge qubits, the minimal times in
the simpler $x$-only control case can be even shorter than in the
$x$-$y$ and $x$-$z$ control cases.  We have also analyzed the
sensitivity of the obtained results for the gate fidelities to random
variations in the control-field amplitudes and finite rise time of the
pulses. Our results are independent of a particular experimental
realization of the $XXZ$ chain, yet, a superconducting Josephson array
would be a particularly appealing candidate.  Our investigation paves
the way for future studies, involving more sophisticated control
strategies~\cite{Makhlin:02,Montangero+:07}.

\begin{acknowledgement}
We would like to thank R. Fazio for discussions.
This work was financially supported by EU project SOLID, the EPSRC
grant EP/F043678/1, the Swiss NSF, and the NCCR Nanoscience.
\end{acknowledgement}

\bibliography{QuantumControl}

\begin{thebibliography}{31}

\bibitem{Lloyd+:04}
S.~Lloyd, A.J. Landahl, J.J.E. Slotine, Phys. Rev. A \textbf{69}, 012305 (2004)

\bibitem{D'AlessandroBook}
D.~D'Alessandro, \emph{Introduction to {Q}uantum {C}ontrol and {D}ynamics}
  (Taylor \& Francis, Boca Raton, 2008)

\bibitem{Jurdjevic+Sussmann:72}
V.~Jurdjevic, H.J. Sussmann, J. Differ. Equations \textbf{12}, 313 (1972)

\bibitem{ControlRevEPJD}
For a recent review, see C. Brif, R. Chakrabarti, H. Rabitz, New. J. Phys.
  ${\mathbf{12}}$, 075008 (2010).

\bibitem{SBose:07}
See, e.g., S. Bose, Phys. Rev. Lett. ${\mathbf{91}}$, 207901 (2003)

\bibitem{Romito+:05}
A.~Romito, R.~Fazio, C.~Bruder, Phys. Rev. B \textbf{71}, 100501(R) (2005)

\bibitem{Lyakhov+Bruder:06}
A.O. Lyakhov, C.~Bruder, Phys. Rev. B \textbf{74}, 235303 (2006)

\bibitem{Burgarth:07}
D.~Burgarth, Eur. Phys. J. Special Topics \textbf{151}, 147 (2007)

\bibitem{Caneva++:09}
T.~Caneva, M.~Murphy, T.~Calarco, R.~Fazio, S.~Montangero, V.~Giovannetti, G.E.
  Santoro, Phys. Rev. Lett. \textbf{103}, 240501 (2009)

\bibitem{Maruyama+:07}
K.~Maruyama, T.~Iitaka, F.~Nori, Phys. Rev. A \textbf{75}, 012325 (2007)

\bibitem{Schirmer:08}
S.G. Schirmer, I.C.H. Pullen, P.J. Pemberton-Ross, Phys. Rev. A \textbf{78},
  062339 (2008)

\bibitem{Burgarth+:09}
D.~Burgarth, S.~Bose, C.~Bruder, V.~Giovannetti, Phys. Rev. A \textbf{79},
  060305(R) (2009)

\bibitem{Kay+Pemberton:10}
A.~Kay, P.J. Pemberton-Ross, Phys. Rev. A \textbf{81}, 010301(R) (2010)

\bibitem{Burgarth+:10}
D.~Burgarth, K.~Maruyama, M.~Murphey, S.~Montangero, T.~Calarco, F.~Nori, M.B.
  Plenio, Phys. Rev. A \textbf{81}, 040303(R) (2010)

\bibitem{Wang+:10}
X.~Wang, A.~Bayat, S.G. Schirmer, S.~Bose, Phys. Rev. A \textbf{81}, 032312
  (2010)

\bibitem{Heule++:10}
R. Heule, C. Bruder, D. Burgarth, V. M. Stojanovi\'c, arXiv:1007.2572 (2010)

\bibitem{Makhlin+:01}
Y.~Makhlin, G.~Sch{\"{o}}n, A.~Shnirman, Rev. Mod. Phys. \textbf{73}, 357
  (2001)

\bibitem{Levitov++:01}
L. S. Levitov, T. P. Orlando, J. B. Majer, J. E. Mooij,
  arXiv:cond-mat/0108266v2 (2001)

\bibitem{You+Nori:05}
J.Q. You, F.~Nori, Phys. Today \textbf{58}, 42 (2005)

\bibitem{GiamarchiBook}
T.~Giamarchi, \emph{{Q}uantum {P}hysics in {O}ne {D}imension} (Clarendon Press,
  Oxford, 2004)

\bibitem{Schirmer+:01}
S.G. Schirmer, H.~Fu, A.I. Solomon, Phys. Rev. A \textbf{63}, 063410 (2001)

\bibitem{PfeiferBook}
W.~Pfeifer, \emph{The {L}ie {A}lgebras $su({N})$: {A}n {I}ntroduction}
  (Birkh{\"{a}}user, Basel, 2003)

\bibitem{Polack+:09}
T.~Polack, H.~Suchowski, D.J. Tannor, Phys. Rev. A \textbf{79}, 053403 (2009)

\bibitem{Sander+Schulte-Herbrueggen:09}
U. Sander, T. Schulte-Herbr\"{u}ggen, arXiv:0904.4654

\bibitem{Burkard++:99}
G. Burkard, D. Loss, D. P. Di Vincenzo, J. A. Smolin, Phys. Rev. B
  ${\mathbf{60}}$, 11404 (1999)

\bibitem{NRfortranBook}
W.H. Press, S.A. Teukolsky, W.T. Vetterling, B.P. Flannery, \emph{Numerical
  {R}ecipes in {F}ortran 77 and 90: {T}he {A}rt of {S}cientific and {P}arallel
  {C}omputing} (Cambridge University Press, Cambridge, 1997)

\bibitem{Carlini+:07}
A.~Carlini, A.~Hosoya, T.~Koike, Y.~Okudaira, Phys. Rev. A \textbf{75}, 042308
  (2007)

\bibitem{bruder++:93}
C.~Bruder, R.~Fazio, G.~Sch{\"{o}}n, Phys. Rev. B \textbf{47}, 342 (1993)

\bibitem{fazio+:01}
R.~Fazio, H.~van~der Zant, Physics Reports \textbf{355}, 235 (2001)

\bibitem{Makhlin:02}
Y.~Makhlin, Quant. Info. Proc. \textbf{1}, 243 (2002)

\bibitem{Montangero+:07}
S.~Montangero, T.~Calarco, R.~Fazio, Phys. Rev. Lett. \textbf{99}, 170501
  (2007)

\end{thebibliography}
\bibliographystyle{epj}
\end{document}